\documentclass[11pt]{article}

\usepackage{deauthor,times,graphicx}

\usepackage{amsmath}
\usepackage{amssymb}

\usepackage{booktabs}
\usepackage{multirow}

\usepackage{tikz}
\usetikzlibrary{positioning, arrows.meta}

\usepackage{hyperref}

\providecommand{\authors}[1]{\author{#1}}

\bulletinyear{2026}

\hypersetup{
  pdftitle    = {Compositional Online Learning for Semantic Data Processing Systems},
  pdfauthor   = {Pawe\l{} Liskowski, Fuheng Zhao, Benjamin Han, Anupam Datta, Dimitris Tsirogiannis},
  pdfkeywords = {LLM-native data systems, query optimization, filter ordering, model cascades, online learning, Cortex AISQL},
}

\begin{document}

\title{Compositional Online Learning for Semantic Data Processing Systems}

\authors{%
  {\normalsize\itshape Pawe\l{} Liskowski \enspace Fuheng Zhao \enspace
   Benjamin Han \enspace Anupam Datta \enspace Dimitris Tsirogiannis}\\[3pt]
  {\normalsize Snowflake Inc.}\\[3pt]
  {\normalsize\texttt{name.surname@snowflake.com}}%
}

\maketitle

\begin{abstract}
An LLM call in a semantic data processing system is expensive enough to dominate query cost, yet slow enough to hide a CPU-side learner's update behind its round-trip. In production, LLM compute accounts for 80--90\% of query cost, and each call costs $10^5$--$10^7\times$ a relational predicate. The latency window inverts a design constraint of classical adaptive query processing, where online learners had to stay lightweight to avoid dominating the predicates they optimize. At LLM latency, per-call gradient steps and per-batch threshold solves fit inside the round-trip.

We develop \emph{compositional online learning at the LLM call boundary}: a framework for combining online-learning components in semantic data processing systems. Each component makes execution-time decisions and refines its learned artifacts online. The design space spans two axes, decision granularity and learner update cadence, and the components share a single learning pattern that hides each trainer step inside the next LLM round-trip.

A production case study in Cortex AISQL composes three components: a memoization layer, an online per-call filter-ordering learner, and an online per-batch cascade-routing learner. A conditional cost decomposition assigns each learning component to a distinct factor of per-row LLM cost. Under independence, the two learning components compose multiplicatively to an $11.4\times$ upper bound on a representative conjunction-filter workload. Self-selection at the cascade boundary, sample-budget shrinkage, and selectivity-estimation drift reduce it to a realistic figure near $8\times$.

\end{abstract}

\section{Introduction}
\label{sec:intro}

Semantic data processing systems extend SQL with operators that invoke a Large Language Model (LLM) on every qualifying row to analyze unstructured data at scale. Industrial systems (Cortex AISQL~\cite{liskowski2025aisql}, BigQuery AI~\cite{HeSethi2025BigQueryAI}, Microsoft Fabric AI Functions~\cite{Microsoft2026FabricAIFunctions}, SEMA-SQL~\cite{lin2026semasqltraditionalrelationalquerying}) and academic systems (LOTUS~\cite{patel2025semantic}, Palimpzest~\cite{liu2025palimpzest}, Quest~\cite{sun2025quest}, ThalamusDB~\cite{Thalamusdb}, SWAN~\cite{zhao2024hybridqueryingrelationaldatabases}) all share the same cost profile, with per-row LLM calls dominating query cost. We propose a framework for optimizing semantic data processing systems through \emph{compositional online learning at the LLM call boundary}. Classical adaptive query processing operates at relational latency, where online learners must be lightweight to avoid dominating the predicates they aim to optimize. At LLM latency, that constraint inverts. Each round-trip is wide enough to hide a CPU-side learner's update, opening a regime where multiple online-learning components can refine the system's per-call decisions during query execution. The resulting design point raises a new question: how do multiple online-learning components compose?

Cortex AISQL, a production deployment of native semantic operators~\cite{liskowski2025aisql}, makes the cost case concrete. A single \texttt{AI\_FILTER} over a million-row table can trigger up to a million LLM calls, each $10^5$--$10^7\times$ more expensive than a relational predicate~\cite{fu2024serverlessllm}. Production data shows that LLM compute accounts for 80--90\% of total query cost, with end-to-end latencies measured in hours rather than seconds~\cite{liskowski2025aisql}. Each call takes hundreds of milliseconds, so a CPU-side learner's update fits inside the round-trip rather than on the critical path.

The composition question is largely open. 
Recent semantic-query systems~\cite{chung2026proxy, dai2024uqe, zhao2025accesspathsefficientordering, liu2025palimpzest, patel2025semantic, Shankaretal2024} optimize individual components offline or at compile time: per-query selectivity samples for filter ordering, per-deployment thresholds for cascade routing, per-pipeline static cost models. Classical adaptive query processing (LEO~\cite{stillger2001leo}, Eddies~\cite{avnur2000eddies}, POLAR~\cite{justen2024polar}, mid-query re-optimization~\cite{kabra1998efficient,markl2004robust}) runs its learners online but at relational latencies, so the techniques transfer in spirit but not in scale. Across both lines of work, components are studied in isolation. To our knowledge, no prior work has analyzed how multiple online-learning components in the LLM-bound regime interact, or whether their gains stack.

We instantiate the framework in Cortex AISQL with two concrete learning components and a non-learning baseline. Filter ordering (Larch~\cite{liskowski2026larch}) refits a per-predicate selectivity model on every LLM outcome and chooses the next predicate per row at online per-call cadence. Cascade routing (GAMCAL~\cite{liskowski2026streamingmodelcascadessemantic}) refits a calibrated GAM on a doubling schedule and routes each row through a cheap proxy or to the oracle at online per-batch cadence. Response caching provides the non-learning baseline by memoizing exact prior calls. We use these three components to develop a composition theory analytically. The two learning components target distinct cost factors and compose multiplicatively under independence. Three cross-component interactions then lower the realistic figure on a representative conjunction-filter workload.

The contributions of this paper are:
\begin{enumerate}
  \item A framework for \textbf{optimizing semantic data processing systems through compositional online learning at the LLM call boundary}, organized around a layered design space (decision granularity $\times$ learner update cadence) and a shared learning pattern that hides each CPU-side learner's update inside the next LLM round-trip, so online per-call gradient steps and online per-batch threshold fits become feasible at scale.

  \item An initial step towards a \textbf{composition theory} for the framework: a sequential composition rule with cost semantics, a conditional cost decomposition that assigns each learning component to a distinct factor of per-row LLM cost, and three named cross-component interactions. The analytical case study yields an $11.4\times$ upper bound under independence and a roughly $8\times$ realistic figure on a representative conjunction-filter workload.
\end{enumerate}


\section{The Framework: Online-Learning Composition at the LLM Call Boundary}
\label{sec:framework}

The LLM round-trip is wide enough to hide a CPU-side online learner's update. That observation drives the framework and has two consequences. First, it admits a class of in-query optimizations that classical adaptive query processing could not afford (\S\ref{sec:aqp-contrast}). Second, it makes one structural pattern repeat across all layers that exploit it (\S\ref{sec:pattern}). The remainder of this section develops the design space the pattern lives in (\S\ref{sec:design-space}), where response caching is the simplest case.

\subsection{Why Classical Adaptive Query Processing Falls Short}
\label{sec:aqp-contrast}

Adaptive query processing was designed for a regime where predicates evaluate in nanoseconds to microseconds. At that budget, online learning has room only for the lightest mechanisms: tuple-level routing as in Eddies~\cite{avnur2000eddies}, plan-shape switches as in mid-query and progressive re-optimization~\cite{kabra1998efficient,markl2004robust,justen2024polar}, or coarse cardinality re-estimation as in feedback-driven optimizers~\cite{stillger2001leo}. Anything more elaborate would dominate the very predicate cost it aims to reduce.

Semantic operators run at a different latency entirely. Per-call LLM latency is hundreds of milliseconds to seconds, which is $10^5$--$10^7\times$ a relational predicate's per-call cost~\cite{liskowski2026larch,fu2024serverlessllm}. Some operators issue many such calls per query: hierarchical text aggregations invoke the LLM at multiple reduction stages, and semantic joins evaluate a predicate on every candidate row pair.

The latency budget defines a new design point. Online per-call gradient updates, online per-batch threshold fits, and combinatorial planning passes over an expression tree all become feasible because anything that fits in a few milliseconds of CPU computation can hide inside the LLM call's round-trip. AQP literature explicitly argues for \emph{lightweight} online learners because relational predicates run in nanoseconds. The latency-window observation inverts that design constraint at LLM latency, where the predicate-cost-dominated optimization rule still holds but online learners can be substantively heavier without violating it. Classical AQP techniques carry over in spirit, but the affordable learner footprint is far larger.

\subsection{Online Learning Behind the LLM Call}
\label{sec:pattern}

One structural pattern recurs across every layer that operates online in the LLM-bound regime. Each LLM call generates one or more supervision signals (the call's output, an oracle judgment on a sampled row, or a ground-truth label). The system buffers them as they arrive, and a learner runs concurrently with the next round's LLM call(s), consuming the buffer and writing back to a learned artifact that the optimizer consults on the next decision.

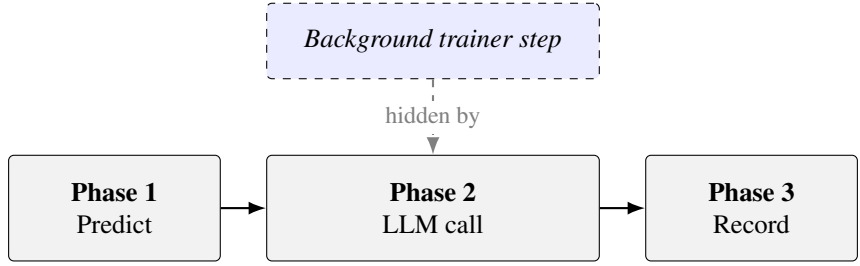
\begin{figure}[t]
  \centering
  \begin{tikzpicture}[
    node distance=6mm,
    phasebox/.style={draw, rounded corners=2pt, align=center,
                     fill=gray!10, minimum height=14mm, font=\small},
    trainerbox/.style={draw, rounded corners=2pt, dashed, align=center,
                       fill=blue!8, minimum height=10mm,
                       font=\small\itshape},
    arr/.style={-Latex, thick},
    parr/.style={-Latex, thick, dashed, gray}]

    \node[phasebox, minimum width=28mm] (predict)
      {\textbf{Phase 1}\\Predict};
    \node[phasebox, minimum width=44mm, right=of predict] (llm)
      {\textbf{Phase 2}\\LLM call};
    \node[phasebox, minimum width=28mm, right=of llm] (record)
      {\textbf{Phase 3}\\Record};
    \node[trainerbox, minimum width=44mm, above=10mm of llm] (trainer)
      {Background trainer step};

    \draw[arr] (predict) -- (llm);
    \draw[arr] (llm) -- (record);
    \draw[parr] (trainer) -- node[font=\footnotesize, fill=white,
                                  inner sep=2pt, midway]{hidden by} (llm);
  \end{tikzpicture}
  \caption{The pipelined online-learning pattern. Phase~1 predicts
    the next decision and dispatches a background trainer step.
    Phase~2 issues the LLM call, which hides the trainer's compute.
    Phase~3 records the outcome into the optimizer's state.
    \S\ref{sec:filter-ordering} instantiates this at per-call cadence
    (one gradient step per LLM call). \S\ref{sec:cascade-routing}
    instantiates it at per-batch cadence (calibration fit and
    threshold solve on a doubling schedule).}
  \label{fig:pipeline}
\end{figure}

Figure~\ref{fig:pipeline} traces the three-phase cycle. In Phase 1, the optimizer queries the current artifact to choose the next decision, and a background thread is dispatched to run a training step on supervision signals buffered from prior rounds. In Phase 2, the chosen evaluation is sent to the LLM, the training step completes inside the round-trip, and the trainer writes back to the artifact. In Phase 3, the LLM's outcome is appended to the supervision buffer for the next trainer step to consume. The trainer thus stays one round behind the supervision but never blocks it.

The pattern is structurally LLM-native: the same updates that would dwarf a relational predicate fit inside the LLM call's round-trip, hidden behind it rather than added to it. In Cortex AISQL, training steps run 7--11\,ms across both variants of Larch~\cite{liskowski2026larch} while LLM evaluations take hundreds of milliseconds, so the trainer's compute fits inside one round-trip. An ablation on the one-round staleness shows that deferring the update keeps per-query token cost within $\pm 0.6\%$ on average across both Larch variants and three datasets. In Larch's deployment, the one-round delay is benign: the training cost is paid in idle CPU cycles, not in the latency budget.

The cadence dictates the affordable update method. At online per-call cadence, the update must fit in one LLM round-trip of typically hundreds of milliseconds, so it stays a lightweight CPU step such as a single gradient. At online per-batch cadence, the update hides behind an entire batch of round-trips ($N$ calls $\times$ per-call latency), so it can be a heavier method such as a Bayesian update or a constrained optimization problem. The two cadences are not interchangeable. Each admits a class of update methods the other cannot afford in practice. Deployment semantics are the same in both cases. The learned artifact is consulted from the first call and refined as new supervision arrives.

Hiding, however, does different work at the two cadences. Online per-call refits lie on the critical path of every LLM call, so a synchronous variant would inflate every call's latency by the trainer's compute. Online per-batch refits fire infrequently (the typical schedule yields $O(\log n)$ events per query), so a synchronous variant would stall only those rare events. Hiding at the online per-batch cadence is a latency-smoothing optimization rather than an online per-call necessity. By contrast, response caching sits at the same per-call boundary but relies on exact memoization, not statistical estimation.

\subsection{The Layered Design Space}
\label{sec:design-space}

Table~\ref{tab:design-space} enumerates the framework's design space along two axes. Each row is a \emph{layer}, a category of execution-time decision the system makes about LLM invocation, and each concrete implementation occupying a layer is a \emph{component}. The framework composes components, and the design space is a typology of the layers components can occupy.

\emph{Decision granularity} asks at what unit each layer decides whether or how the LLM is invoked: per-call, per-row, or per-batch in the LLM-bound regime considered here. Other granularities (per-tuple, per-pipeline, per-query) appear in classical AQP. \emph{Learner update cadence} asks when the layer's learned artifact is refined relative to query execution: online (per-call or per-batch), one-shot at compile time, offline, or none (memoization). Response caching, filter ordering, and cascade routing are the three layers the AISQL instantiation fills.

\begin{table}[t]
  \centering
  \caption{The layered design space. Three rows are populated by
    the AISQL instantiation: response caching (\S\ref{sec:caching}),
    filter ordering (\S\ref{sec:filter-ordering}), and cascade routing
    (\S\ref{sec:cascade-routing}). The remaining four rows name
    LLM-bound research directions revisited in
    \S\ref{sec:open-layers}. AQP-style layers populate analogous
    decisions at relational latency and are discussed in
    \S\ref{sec:related-work} rather than in this LLM-bound design space.}
  \label{tab:design-space}
  \resizebox{\textwidth}{!}{%
  \begin{tabular}{@{}l l l l l@{}}
    \toprule
    Layer & Decision granularity & Learner update cadence & Artifact / state & Instance \\
    \midrule
    Response caching   & per-call & none (memoization)
                       & exact-match response cache
                       & AISQL response cache (\S\ref{sec:caching}) \\
    Filter ordering    & per-call & online per-call
                       & selectivity model + DP solver
                       & \textbf{Larch} (\S\ref{sec:filter-ordering}) \\
    Cascade routing    & per-row  & online per-batch
                       & calibrated GAM probability map
                       & \textbf{GAMCAL} (\S\ref{sec:cascade-routing}) \\
    \midrule
    Prompt caching     & per-call & none (memoization)
                       & prefix cache + affinity map
                       & future direction (\S\ref{sec:open-layers}) \\
    Model substitution & per-batch & one-shot at compile time
                       & distilled student net
                       & future direction (\S\ref{sec:open-layers}) \\
    Prompt structure   & per-call & online per-batch
                       & prompt template / rewriting rules
                       & future direction (\S\ref{sec:open-layers}) \\
    Adaptive batching  & per-batch & online per-batch
                       & batch-size schedule
                       & future direction (\S\ref{sec:open-layers}) \\
    \bottomrule
  \end{tabular}%
  }
\end{table}

The AISQL instantiation exercises both axes. Filter ordering (Larch, \S\ref{sec:filter-ordering}) runs at online per-call cadence and targets $\bar{e}$, the expected number of LLM-evaluated predicates per row, through short-circuit ordering. Cascade routing (GAMCAL, \S\ref{sec:cascade-routing}) runs at online per-batch cadence and targets $p_{\text{deleg}}$, the fraction of evaluated predicates sent past the proxy to the oracle, through calibrated thresholds. Both cost factors are formalized in \S\ref{sec:cost-decomp}.

The two components compose multiplicatively under independence. \S\ref{sec:composition} formalizes this as a sequential rule with cost semantics, decomposes the per-row LLM cost across the two components, and catalogs the cross-component interactions that set the realistic figure. \phantomsection\label{sec:caching}Response caching, the third such cell, has no learner or update cadence: a hit on an exact-match key returns the stored response and suppresses both learning components, so savings track the duplicate fraction. Because hit rates can correlate with predicate position or row subpopulation, the cost model conditions on the non-cached path instead of parameterizing cache effects.

\section{The Cortex AISQL Substrate}
\label{sec:substrate}

Cortex AISQL is a production system that embeds native semantic operators in relational SQL and the case study for this framework~\cite{liskowski2025aisql}. It provides six operators: \texttt{AI\_FILTER} for per-row boolean predicates, \texttt{AI\_CLASSIFY} for per-row labels from a fixed set, \texttt{AI\_COMPLETE} for text generation, \texttt{AI\_AGG} and \texttt{AI\_SUMMARIZE\_AGG} for text-column aggregations, and \texttt{AI\_JOIN} for semantic joins over row pairs. Each invokes an LLM on every qualifying row or row pair and composes with relational operators in one declarative statement. The two learning components target \texttt{AI\_FILTER}, so the cost decomposition in \S\ref{sec:cost-decomp} is scoped to that operator and generalizes to any per-row semantic predicate.

To the optimizer, these operators are black boxes whose cost and selectivity are both unknown at compile time. Per-row cost depends on the model, prompt length, and inference infrastructure, while selectivity depends on a natural-language predicate with no histogram to read against. AISQL's existing optimizer addresses both unknowns in two stages. At compile time, cost-aware placement pulls expensive \texttt{AI\_FILTER} predicates above joins and orders semantic predicates by relative cost. At runtime, coarse per-predicate cost and selectivity statistics trigger a reorder when a different order proves more effective on the observed batch. The framework's two learning components extend this runtime side, replacing passively collected population statistics with per-outcome online refinement.
\section{Filter Ordering}
\label{sec:filter-ordering}

Because AND/OR are commutative, the evaluation order of an \texttt{AI\_FILTER} conjunction or disjunction changes only the cost, never the row's truth value. Order determines how often short-circuit reduction prunes the remaining predicates, and the optimal order at each row depends on per-row pass probabilities that classical optimizers cannot pre-compute. Larch instantiates the filter-ordering component on AISQL \texttt{AI\_FILTER} conjunctions. It refits a per-predicate selectivity model after every LLM outcome and reorders the remaining predicates per row, both at online per-call cadence.

Larch-Sel decomposes the problem into two pieces, an online per-predicate selectivity estimator and an exact dynamic-programming solver over the AND/OR tree. The estimator is a two-layer MLP with roughly 144K trainable parameters, sigmoid output, and binary cross-entropy training. Its input is built from pre-computed document and predicate embeddings, produced at ingestion and exposed at query time at a cost roughly two orders of magnitude below an \texttt{AI\_FILTER} evaluation~\cite{liskowski2026larch}. Each embedding is first projected to 64 dimensions via learned linear maps $\mathbf{W}_{\mathrm{doc}}$ and $\mathbf{W}_{\mathrm{filter}}$, then combined into a feature vector that captures per-side identity, multiplicative interaction, and directional alignment:
\begin{equation*}
  \mathbf{x} = [\,\mathbf{d} \,\|\, \mathbf{f} \,\|\, \mathbf{d} \odot \mathbf{f} \,\|\, \cos(\mathbf{d}, \mathbf{f})\,].
\end{equation*}
All predicates share the MLP's weights, so the model transfers what it has learned about one predicate to the next from the first evaluation onward.

The DP solver evaluates the recurrence
\begin{equation}
  \mathrm{OPT}(T') =
  \begin{cases}
    0 & T' \text{ resolved},\\
    \min\limits_{f_i \in \mathrm{leaves}(T')}\!\bigl[\,c_i + \hat{s}_i \cdot \mathrm{OPT}(T'|_{f_i=\top}) + (1 - \hat{s}_i) \cdot \mathrm{OPT}(T'|_{f_i=\bot})\,\bigr] & \text{otherwise,}
  \end{cases}
  \label{eq:opt}
\end{equation}
where $\mathrm{OPT}(T')$ is the minimum expected cost to resolve a partially evaluated tree $T'$ given the current selectivity estimates. Solving the recurrence for one document runs in $O(n \cdot 3^n)$ time ($\approx 590$K operations at $n{=}10$, $\approx 20$\,ms on a single CPU core). A sibling end-to-end RL variant, Larch-A2C, encodes the expression tree with a Gated Graph Neural Network and is the appropriate fit when predicate outcomes within a row are correlated.

The decomposition exploits a structural property: if per-predicate pass probabilities are accurate, the minimum-cost evaluation order can be solved exactly under the independence assumption. The DP recurrence (Eq.~\ref{eq:opt}) generalizes Krishnamurthy, Boral, and Zaniolo's heterogeneous-cost ordering theorem~\cite{krishnamurthy1986optimization} from flat conjunctions to arbitrary AND/OR trees. The original theorem itself builds on Ibaraki and Kameda's uniform-cost result~\cite{ibaraki1984optimal}. Accurate per-row selectivities therefore suffice for an exactly-optimal ordering, leaving no residual signal for a learned policy to extract. Filter ordering is \emph{estimation-dominated}, so a sample-efficient selectivity model matters more than a richer planner. Empirically, Larch-Sel reaches near-optimal performance from a few hundred rows on the smallest benchmark, while Larch-A2C closes the gap to the lower bound only as the horizon grows~\cite{liskowski2026larch}.

Larch-Sel instantiates the framework's learning pattern. Each \texttt{AI\_FILTER} outcome is a binary label that supervises one BCE gradient step on the shared selectivity MLP. The step runs in a background thread during the next predicate's LLM call and completes in roughly 7\,ms on CPU, inside the LLM's hundreds-of-millisecond round-trip. The pipeline is one step stale, with training during round $t{+}1$ using observations from round $t$. The asynchronous structure introduces one cross-component interaction with cascade routing.

Larch-Sel reduces total token-cost overhead by up to $19\times$ relative to Palimpzest (PZ)~\cite{liu2025palimpzest} and Quest~\cite{sun2025quest}, with typical Mix-workload reductions in the 4--8$\times$ band~\cite{liskowski2026larch}. Table~\ref{tab:overhead} shows the per-dataset slice. The reduction comes from per-document selectivity carrying strictly more information than the population averages PZ and Quest learn from. The extra signal lets Larch-Sel match or exceed an oracle baseline with access to ground-truth global selectivities. PZ and Quest also pay an upfront 5\% sampling cost in actual \texttt{AI\_FILTER} calls to estimate those averages before execution begins. Larch-Sel pays no startup tax: it learns from the same \texttt{AI\_FILTER} calls the query already performs.

\begin{table}[t]
  \centering
  \caption{Token-overhead percentages relative to the per-row
    optimal lower bound on the Mix workload (50\% AND, 50\% OR)
    across three datasets~\cite{liskowski2026larch}. Lower is
    better.}
  \label{tab:overhead}
  \begin{tabular}{l c c c}
    \toprule
    Dataset   & PZ      & Quest   & \textbf{Larch-Sel}    \\
    \midrule
    GovReport & 21.3\% & 33.6\% & \textbf{5.1\%}        \\
    PubMed    & 26.1\% & 29.5\% & \textbf{6.5\%}        \\
    BigPatent & 28.3\% & 29.7\% & \textbf{3.6\%}        \\
    \bottomrule
  \end{tabular}
\end{table}

Larch reduces $\bar{e}$, the expected predicate evaluations per row, through short-circuit ordering. Prior filter ordering instead fixes its estimates before execution, one-shot or offline (Palimpzest, Quest, SWAN, SEMA-SQL), as \S\ref{sec:related-work} details.

\section{Cascade Routing}
\label{sec:cascade-routing}

In cascade routing, we use two proxy score thresholds to partition the rows: rows below $\tau_{\text{low}}$ are rejected, rows at or above $\tau_{\text{high}}$ are accepted, and rows in the uncertain region $[\tau_{\text{low}}, \tau_{\text{high}})$ are delegated to the oracle LLM. AISQL's production cascade already pairs the two models. The proxy runs on every row at low cost, while the oracle accounts for the dominant per-call cost. Cost minimization reduces to choosing the threshold pair that balances classification quality against the oracle delegation rate. GAMCAL instantiates the cascade-routing component on AISQL \texttt{AI\_FILTER} rows, refitting a calibrated GAM and re-solving the thresholds as oracle labels accumulate across batches, at online per-batch cadence~\cite{liskowski2026streamingmodelcascadessemantic}.

GAMCAL fits a Generalized Additive Model (GAM)~\cite{hastie1986generalized,wood2017generalized} that maps raw proxy scores to calibrated true-positive probabilities. Thresholds are then chosen to minimize a single cost-quality objective:
\begin{equation}
  \min_{\tau_{\text{low}} \leq \tau_{\text{high}}}\;
  \alpha \cdot \frac{1 - \mathbb{E}[F_1(\tau_{\text{low}},
    \tau_{\text{high}})]}{1 - \mathbb{E}[F_1(0.5, 0.5)]}
  + (1 - \alpha) \cdot
  \mathrm{deleg}(\tau_{\text{low}}, \tau_{\text{high}}),
  \label{eq:cost-quality}
\end{equation}
where $\alpha \in [0, 1]$ is a user-controlled trade-off knob, the first term is the normalized $F_1$ degradation against the no-delegation baseline, and $\mathrm{deleg}$ is the delegation rate. The calibrator $f$ is a smoothing spline in log-odds space, fit on the accumulated oracle-label buffer $S$ by penalized maximum likelihood with a smoothness penalty and a monotonicity constraint:
\begin{equation}
  f^{\star} = \arg\max_{f \text{ monotone}}
  \sum_{(s, y) \in S}\!\bigl[y \log \sigma(f(s)) + (1{-}y) \log(1 - \sigma(f(s)))\bigr]
  - \lambda \!\int_0^1\! \bigl(f''(s)\bigr)^2 \, ds.
  \label{eq:gam-fit}
\end{equation}

GAMCAL routes each row through a stochastic calibrated score, computed as the GAM's mean prediction plus a Gaussian draw scaled by the calibrator's posterior standard error:
\begin{equation}
  \tilde{g}_i = \sigma\!\bigl(f(s_i) + \Phi^{-1}(q_i) \cdot \mathrm{se}(s_i)\bigr),
  \label{eq:calibrated-score}
\end{equation}
with $s_i$ the row's proxy score, $q_i \sim U(0, 1)$ a per-row quantile drawn once and held fixed for the row's lifetime, $\sigma$ the logistic, and $\Phi^{-1}$ the standard-normal quantile function. Rows in poorly calibrated regions of the score range receive more dispersed calibrated scores and tend to land in the uncertain region, so routing explores without an explicit importance-sampling scheme. Both calibrator and thresholds refit on a doubling schedule, holding retraining events to $O(\log n)$ per query. Each batch $B_t$ adds $\rho |B_t|$ uniformly sampled rows from the uncertain region to a labeled buffer, and the first refit fires once both classes contain at least $n_{\min}$ labels.

A sibling target-based algorithm, SUPG-IT, handles workloads with explicit precision-recall contracts. We focus on GAMCAL because its smooth $\alpha$ knob composes directly with the filter-ordering component's cost lever. Two arguments motivate learned calibration over the SUPG family of statistical threshold estimators. First, modern proxy classifiers are often poorly calibrated~\cite{guo2017calibration}. Importance-weighted thresholding methods that assume calibrated proxies therefore produce conservative thresholds with high delegation rates. Second, the GAM generalizes oracle labels across the full proxy-score distribution. The spline interpolates each label at one score level into predictions at every other level, while running-statistics estimators in the SUPG family~\cite{kang2020approximate} extract information only from the score region around each observed label.

GAMCAL instantiates the framework's learning pattern at batch scale (Figure~\ref{fig:pipeline}). Each refit runs concurrently with the spawning batch's oracle calls and is installed as the routing state at the start of the next batch. At a batch size of roughly 4{,}000 rows with oracle calls taking hundreds of milliseconds, the aggregate per-batch oracle window absorbs the CPU GAM fit and threshold solve. The pipeline lags by one batch: routing in batch $t{+}1$ uses calibration fit from labels through batch $t{-}1$. The lag mirrors at batch scale what filter ordering sees at call scale and introduces another cross-component interaction.

GAMCAL exceeds $F_1 = 0.95$ at its best operating point on every one of six benchmarks spanning classification, filtering, and join workloads~\cite{liskowski2026streamingmodelcascadessemantic}. To reach that threshold against LOTUS's single-pass SUPG cascade~\cite{patel2025semantic}, GAMCAL needs up to $58\%$ fewer oracle calls and leads on five of six benchmarks (Table~\ref{tab:cascade-overhead}). ArXiv is the exception: its proxy is already well-calibrated, so importance-weighted thresholding suffices and there is little for the GAM to recover. The advantage is largest on BoolQ and MMLU, where the proxy is reasonably accurate but poorly calibrated, so the GAM's smooth interpolation gains a clear sample-efficiency edge over running statistics.

\begin{table}[t]
  \centering
  \caption{Minimum delegation rate $d$ required to reach
    $F_1 \geq 0.95$ across all six
    benchmarks~\cite{liskowski2026streamingmodelcascadessemantic}.
    Lower is better. Bold marks the best algorithm per row.
    GAMCAL leads on five of six datasets, often by a wide margin.}
  \label{tab:cascade-overhead}
  \begin{tabular}{l c c c}
    \toprule
    Dataset & SUPG-SP (LOTUS) & SUPG-IT & GAMCAL \\
    \midrule
    ArXiv & \textbf{55.6\%} & 61.9\% & 61.6\% \\
    BoolQ & 69.7\% & 58.2\% & \textbf{29.3\%} \\
    IMDB  & 69.9\% & 69.1\% & \textbf{68.4\%} \\
    MMLU  & 67.4\% & 62.9\% & \textbf{43.9\%} \\
    NYT   & 21.9\% & 22.4\% & \textbf{17.0\%} \\
    SST-2 & 28.6\% & 31.6\% & \textbf{21.2\%} \\
    \bottomrule
  \end{tabular}
\end{table}

GAMCAL's calibrated thresholds reduce $p_{\text{deleg}}$, the fraction of evaluated predicates delegated past the proxy to the oracle and the second factor of the cost decomposition. \S\ref{sec:composition} composes this reduction with Larch's into the multiplicative bound, and \S\ref{sec:related-work} sets GAMCAL's per-batch calibration against the offline and one-shot cascades (FrugalGPT, SUPG, LOTUS).

\section{Composition Across Learning Components}
\label{sec:composition}

\subsection{Sequential Composition with Cost Semantics}
\label{sec:composition-rule}

The two learning components, filter ordering (Larch) and cascade routing (GAMCAL), run in a fixed order: filter ordering first, then per-row cascade routing on each evaluated predicate. Total cost depends on each component's standalone behavior and on how the upstream reshapes the distribution the downstream sees, so we state composition as a Hoare-style rule with cost semantics. The rule makes the multiplicative-decomposition assumption explicit and classifies cross-component interactions into forward precondition mismatches and a backward supervision flow.

Following the standard form of Hoare's sequential composition rule and its probabilistic extensions with cost semantics~\cite{hoare1969axiomatic, kaminski2018weakest}, we write each learning component as a triple
\begin{equation*}
  \langle D_{\text{in}},\, S,\, D_{\text{out}} \rangle_{c(D_{\text{in}})},
\end{equation*}
where the precondition $D_{\text{in}}$ is the input distribution over rows or (row, predicate) pairs, the program $S$ is the component's online learner plus decision rule, the postcondition $D_{\text{out}}$ is the distribution of items the component forwards to the next component, and $c(D_{\text{in}})$ is the expected per-input LLM cost the component pays as a function of the input distribution. Parameterizing cost by $D_{\text{in}}$ rather than a nominal value matters because Larch and GAMCAL train against the rows they actually see, so their realized cost shifts with the distribution the upstream produces.

Sequential composition combines two triples into one. The downstream's cost is weighted by the upstream's fan-out $\phi_1(D_0)$, the expected number of items $S_1$ forwards per input:
\begin{equation}
  \frac{
    \langle D_0,\, S_1,\, D_1 \rangle_{c_1(D_0)}
    \quad
    \langle D_1,\, S_2,\, D_2 \rangle_{c_2(D_1)}
  }{
    \langle D_0,\, S_1; S_2,\, D_2 \rangle_{c_1(D_0)\, +\,
      \phi_1(D_0) \cdot c_2(D_1)}
  }.
  \label{eq:seq-composition}
\end{equation}
Composition is sound only when $S_2$'s precondition is the same distribution $S_1$ produces, mirroring Hoare logic's postcondition-precondition matching.

For the Larch + GAMCAL composition, Larch ($S_1$) makes no LLM calls of its own ($c_1 \equiv 0$), forwards $\bar{e}$ evaluated predicates per row in expectation ($\phi_1(D_0) = \bar{e}$), and produces a postcondition $D_1$ over (row, predicate) pairs sent to the cascade. GAMCAL ($S_2$) pays per-evaluated-predicate cost $c_2(D_1) = c_{\text{proxy}} + p_{\text{deleg}}(D_1) \cdot c_{\text{oracle}}$, with $p_{\text{deleg}}$ the delegation rate evaluated on Larch's actual postcondition. Substituting into Eq.~\ref{eq:seq-composition} yields the per-row form
\begin{equation}
  \bar{e} \cdot \bigl(c_{\text{proxy}} + p_{\text{deleg}}(D_1) \cdot
    c_{\text{oracle}}\bigr).
  \label{eq:per-row-form}
\end{equation}
When $D_1$ matches GAMCAL's calibration distribution $D_1^\star$, $p_{\text{deleg}}(D_1)$ takes its nominal calibrated value, and Eq.~\ref{eq:per-row-form} is the matched-precondition instance of Eq.~\ref{eq:seq-composition}. \S\ref{sec:cost-decomp} states the workload assumptions behind the form.

The per-row form depends only on the realized values of $\bar{e}$ and $p_{\text{deleg}}(D_1)$, not on the runtime sequence in which Larch chooses predicates and the cascade evaluates them. Cross-component interactions are the mechanisms that shift those values. When $D_1 \neq D_1^\star$, GAMCAL's per-item cost picks up a multiplicative penalty $c_2(D_1) / c_2(D_1^\star)$, and the composition rule classifies the three by direction. Self-selection at the cascade boundary ($\kappa_1$) and sample-budget shrinkage ($\kappa_2$) are forward precondition mismatches, while selectivity-estimation drift ($\kappa_3$) is a backward supervision flow from cascade outputs into Larch's selectivity learner.

The two-component rule extends by induction to longer chains of learning components. Each component whose cost is parameterized by its input distribution fits the triple notation, and the composed per-input cost decomposes by the same fan-out-weighted accounting. Two extensions lie outside the present formulation: branching topologies, where one component fans out to multiple downstreams, and backward supervision flows like $\kappa_3$, where downstream outputs return as labels into an upstream learner. Both are future work.

\subsection{Conditional Cost Decomposition}
\label{sec:cost-decomp}

For the conjunction-filter workload, the per-row form (Eq.~\ref{eq:per-row-form}) becomes an explicit cost equation once its four parameters and underlying assumptions are stated. The natural unit of analysis is \emph{expected predicate evaluations} per row ($\bar{e}$), because an \texttt{AI\_FILTER} conjunction with $m$ predicates can incur up to $m$ LLM calls per row in the unoptimized case.

For $N$ rows flowing through a filter conjunction with $m$ semantic predicates and per-predicate cascade routing, total cost factors as
\begin{equation}
  \mathrm{Cost} = N \cdot \bar{e} \cdot
    \big[\, c_{\text{proxy}} + p_{\text{deleg}} \cdot c_{\text{oracle}} \,\big],
  \label{eq:cost-decomp}
\end{equation}
with the four parameters defined as follows:
\begin{itemize}
  \item $\bar{e} \in [1, m]$, the expected number of LLM-evaluated
    predicates per row on the non-trivial conjunction path, equal to
    $m$ in the unoptimized baseline.
  \item $c_{\text{proxy}}$, the per-call proxy cost, equal to
    $0$ in the no-cascade baseline.
  \item $p_{\text{deleg}} \in [0, 1]$, the per-predicate oracle
    delegation rate, equal to $1$ in the no-cascade baseline.
  \item $c_{\text{oracle}}$, the per-call oracle cost, which
    dominates the unoptimized total.
\end{itemize}

Each learning component targets a distinct factor of Eq.~\ref{eq:cost-decomp}. Larch reduces $\bar{e}$ via short-circuit ordering. The cascade reduces $p_{\text{deleg}}$ via calibrated thresholds at the cost of paying $c_{\text{proxy}}$ on every evaluated predicate. Under independence, the two reductions compose multiplicatively, the matched-precondition case of the sequential composition rule (\S\ref{sec:composition-rule}).

The decomposition rests on three assumptions: per-call costs are uniform across rows and predicates, predicate selectivities are independent within a row, and the analysis is conditioned on a cache miss. The cache-miss conditioning means Eq.~\ref{eq:cost-decomp} describes the non-cached path, with cache hits held outside the model rather than parameterized within it. The decomposition extends to non-conjunctive expression trees by generalizing $\bar{e}$ through the DP recurrence (Eq.~\ref{eq:opt}) over AND/OR nodes.

The independence and cache-miss assumptions each hide a workload interaction. Within-row independence fails because rows that survived earlier predicates have correlated pass probabilities downstream, and the self-selection interaction $\kappa_1$ captures the resulting shift in GAMCAL's input distribution. A joint cache-and-learning model would require an analogous workload-specific assumption, since cache hit rates vary with predicate position, row subpopulation, and routing decisions. Quantifying that variation requires evidence beyond what this paper presents, so cache effects remain outside the displayed decomposition.

\subsection{An Analytical Case Study}
\label{sec:case-study}

A representative conjunction-filter workload has $N = 10^6$ rows, $m = 5$ semantic predicates in the WHERE clause, and proxy-to-oracle cost ratio $c_{\text{proxy}} / c_{\text{oracle}} \approx 0.05$. Each row of Table~\ref{tab:composition} substitutes one configuration's parameter values into Eq.~\ref{eq:cost-decomp} and reports the resulting cost relative to the no-optimization baseline.

\begin{table}[t]
  \centering
  \caption{Cost relative to the no-optimization baseline at $m = 5$
    and $c_{\text{proxy}} / c_{\text{oracle}} = 0.05$. The
    two-component upper bound is $11.4\times$ under
    \S\ref{sec:cost-decomp}'s independence assumption and matched-precondition
    calibration ($D_1 = D_1^\star$ in the sense of
    \S\ref{sec:composition-rule}).
    \S\ref{sec:interactions} catalogs cross-component interactions
    that lower the realistic figure to roughly $8\times$.}
  \label{tab:composition}
  \begin{tabular}{l c c c}
    \toprule
    Configuration                   & $\bar{e}$        & $p_{\text{deleg}}$ & Cost (rel.) \\
    \midrule
    Baseline (no optimization)      & 5.0              & 1.0                & 1.00              \\
    + Filter ordering alone         & $\approx 1.25$   & 1.0                & $\approx 0.25$    \\
    + Cascade routing alone         & 5.0              & $\approx 0.30$     & $\approx 0.35$    \\
    \textbf{Both learning components (upper bound)} & $\approx 1.25$   & $\approx 0.30$     & $\approx \mathbf{0.0875}$ ($11.4\times$) \\
    \bottomrule
  \end{tabular}
\end{table}

$\bar{e} \approx 1.25$ at $m = 5$ corresponds to a $4\times$ cut in expected predicate evaluations, at the lower end of Larch-Sel's typical $4$--$8\times$ Mix-workload band and well below its $19\times$ ceiling against Palimpzest and Quest~\cite{liskowski2026larch}. Counting LLM calls instead of token cost keeps $\bar{e} \geq 1$, since every row evaluates at least one predicate to resolve the conjunction. $p_{\text{deleg}} \approx 0.30$ falls inside GAMCAL's $17$--$68\%$ range of minimum delegation rates to reach $F_1 \geq 0.95$ across six benchmarks~\cite{liskowski2026streamingmodelcascadessemantic}, near the easier-to-cascade end. The cost ratio $c_{\text{proxy}} / c_{\text{oracle}} \approx 0.05$ follows from inference-cost scaling: compute scales with model parameters at roughly $10\times$ between an 8B proxy and a 70B oracle, and drops further when the smaller model runs on cheaper hardware~\cite{fu2024serverlessllm}.

The final row substitutes the two optimized values into Eq.~\ref{eq:cost-decomp}:
\begin{equation*}
  \mathrm{Cost}_{\text{both}}
    = N \cdot 1.25 \cdot \bigl[\, 0.05 \cdot c_{\text{oracle}}
                                  + 0.30 \cdot c_{\text{oracle}} \,\bigr]
    = N \cdot 0.4375 \cdot c_{\text{oracle}}
    \approx 0.0875 \cdot \mathrm{baseline}.
\end{equation*}
The naive product of the two standalone cost ratios, $0.25 \times 0.35 = 0.0875$, matches the formula exactly because $c_{\text{proxy}}$ enters the per-call cost factor that both components multiply through. The $11.4\times$ figure is therefore the upper bound under independence and matched-precondition calibration ($D_1 = D_1^\star$). Three cross-component interactions push the realistic figure to roughly $8\times$.

The $11.4\times$ figure holds approximately constant for $m \geq 4$, where Larch's reduction does not hit the $\bar{e} \geq 1$ floor. Cost ratio and $p_{\text{deleg}}$ matter more: a cost ratio of $0.10$ compresses it to about $10\times$, and the harder end of GAMCAL's range ($p_{\text{deleg}} \approx 0.68$) to about $5.5\times$.

\subsection{Cross-Component Interactions}
\label{sec:interactions}

Three interactions lower the $11.4\times$ upper bound, each a multiplicative penalty $\kappa_i$ on one term of Eq.~\ref{eq:cost-decomp}. We treat the three as independent. Mechanistic coupling between $\kappa_1$ and $\kappa_3$ would push the realistic figure below $8\times$.

\textbf{Self-selection at the cascade boundary ($\kappa_1$).} Larch routes only conjunction-surviving rows to a given cascade-evaluated predicate, and survivors skew toward harder cases with proxy scores nearer the decision boundary. The per-batch refresh tracks shift between batches, but steady-state $p_{\text{deleg}}$ on this self-selected stream stays above the global rate. We bound the penalty at $\kappa_1 \in [1.0, 1.5]$ on $p_{\text{deleg}}$. At the high end, delegation rises by roughly $50\%$ over the global rate.

\textbf{Sample-budget shrinkage ($\kappa_2$).} The cascade draws its $\rho$-fraction sample budget per batch, so Larch's short-circuiting shrinks the effective batch at later predicates and yields fewer oracle labels per refinement there, slowing convergence. Because the shortfall peaks before thresholds stabilize, the penalty $\kappa_2 \in [1.0, 1.3]$ falls on early-query $p_{\text{deleg}}$ alone, its upper end a cascade running at about $70\%$ of steady-state effectiveness during cold start.

\textbf{Selectivity-estimation drift across the cascade ($\kappa_3$).} Unlike the forward mismatches $\kappa_1$ and $\kappa_2$, this one runs backward. Larch's selectivity MLP trains on predicate outcomes, but when a downstream filter is served by a cascade that returns proxy decisions in its accept and reject regions, those labels are no longer pure oracle labels and the gradient signal grows noisier in proportion to the accept-and-reject fraction. The drift puts $\kappa_3 \in [1.0, 1.4]$ on $\bar{e}$, reaching its maximum at an accept-and-reject fraction near $70\%$, where degraded selectivity precision yields less aggressive short-circuiting.

$\kappa_2$ contributes during the cascade's pre-steady-state phase only, so its contribution averages out over a long operator. On a $10^6$-row query at batch size $4{,}096$ and $\rho = 0.10$, the pre-steady-state phase covers roughly the first $1\%$ of batches under the doubling schedule. The operator-level sweep holds $\kappa_2 = 1$ and combines $\kappa_1$ on $p_{\text{deleg}}$ with $\kappa_3$ on $\bar{e}$, giving:
\begin{equation}
  \mathrm{Cost}_{\text{rel}} = 0.25 \cdot \kappa_3 \cdot (0.05 + 0.30 \cdot \kappa_1).
  \label{eq:realistic-cost}
\end{equation}
At midpoint values $\kappa_1 = 1.25$ and $\kappa_3 = 1.2$, this yields $\sim 7.8\times$, which we round to $8\times$ as the realistic figure. The pessimistic corner ($\kappa_1 = 1.5$, $\kappa_3 = 1.4$) yields $\sim 5.7\times$, and the optimistic corner ($\kappa_1 = \kappa_3 = 1$) recovers the $11.4\times$ upper bound exactly.


\section{Related Work}
\label{sec:related-work}

On the framework's two axes (decision granularity, learner update cadence), the AISQL instantiation is the only system to fill the \emph{online per-call} and \emph{online per-batch} cells in the LLM-bound regime. Other LLM-bound systems operate at the same decision granularities but at one-shot or offline cadences, and classical AQP runs online only at relational latency, where the latency-window observation does not apply. At LLM latency, a constraint the AQP and cascade literatures treated as fixed no longer holds: one-shot or offline cadences become optional rather than necessary.

\textbf{AQP at relational latency.} Classical AQP spans per-tuple, per-pipeline, and per-query granularities at online cadences (Eddies~\cite{avnur2000eddies}; POLAR and mid-query reopt~\cite{justen2024polar,kabra1998efficient,markl2004robust}; LEO~\cite{stillger2001leo} and ThalamusDB~\cite{Thalamusdb}), but only at relational latency, where the predicate-cost constraint analyzed in \S\ref{sec:aqp-contrast} forces sub-microsecond learners. The latency window does not apply there, so these techniques apply conceptually but at a far smaller learner footprint.

\textbf{The filter-ordering cadence argument.} Prior LLM-aware filter ordering fixes its selectivity estimates once, before execution. Palimpzest~\cite{liu2025palimpzest}, Quest~\cite{sun2025quest}, SWAN~\cite{zhao2024hybridqueryingrelationaldatabases}, and SEMA-SQL~\cite{lin2026semasqltraditionalrelationalquerying} sit in the \emph{one-shot at compile time} cell, learning population-level averages from a sampling pass and freezing them. Larch instead refines a per-document selectivity model on every LLM outcome, occupying the \emph{online per-call} cell that prior work leaves empty. Model substitution shows the same pattern at coarser granularity: UQE~\cite{dai2024uqe} distills a student model at compile time rather than online.

\textbf{The cascade cadence argument.} Prior cascade work populates the cascade-routing layer at two cadences, offline and one-shot at compile time. SUPG~\cite{kang2020approximate}, FrugalGPT~\cite{chen2023frugalgpt}, and BigQuery proxy work~\cite{chung2026proxy} fit calibration sets or proxy classifiers before deployment and freeze them afterward. LOTUS~\cite{patel2025semantic} deploys a variant called SUPG-SP that fits its threshold on a per-query calibration sample at compile time, placing it in the \emph{one-shot at compile time} cell. GAMCAL~\cite{liskowski2026streamingmodelcascadessemantic} extends both to an \emph{online per-batch} cadence in which calibration is refreshed during query execution on a doubling schedule. The framework's distinctive contribution to cascade theory is the online per-batch cadence, made feasible by the latency window.

The online cadences the latency window opens thus remain largely empty in practice: across LLM-aware filter ordering, cascade routing, and model substitution, only Larch and GAMCAL sit at online cadences, while the rest occupy compile-time or offline cells consistent with the AQP inversion argument.

\section{Discussion}
\label{sec:discussion}

The framework and its composition analysis leave four open directions: filling the rest of the design surface, tightening composition across the components in place, extending the learning loop across query boundaries, and reconciling the components' heterogeneous quality semantics.

\subsection{Future Layers in the Design Space}
\label{sec:open-layers}

Response caching shows that a productive layer can rely on memoization rather than online estimation; the remaining design-space rows stay open. \emph{Prompt caching} shares that mechanism, but its hit rates correlate with predicate position and routing in ways the displayed cost decomposition does not parameterize. \emph{Model substitution} generalizes the cascade to per-batch granularity: UQE's distilled student~\cite{dai2024uqe} sits at the one-shot cell, and the latency window admits an online per-batch variant that refits both student and gate. \emph{Prompt structure} shapes the call itself; AISQL's join-to-classify rewrite~\cite{liskowski2025aisql} is a compile-time instance, while an online variant could key rewrite rules to observed outcomes, dropping error-correlated exemplars and restoring hallucination-correlated hints. \emph{Adaptive batching} trades per-call latency, throughput, and quality, and fits the online per-batch cadence directly. These layers couple through shared prompt, model, and batch choices, so their refit signals belong in a joint controller rather than in isolated components.

\textbf{Joint optimization across components.} The two learning components are tuned independently today, but the interactions suggest concrete couplings a controller could exploit: refit the cascade against the post-Larch self-selected stream ($\kappa_1$), size its sample budget to the post-short-circuit batch at each predicate position ($\kappa_2$), and down-weight Larch's gradient signal from labels collected in the cascade's accept and reject regions ($\kappa_3$). The natural interface generalizes GAMCAL's single-$\alpha$ knob across both components, exposing the composed system's cost-quality frontier rather than each component's in isolation. The same controller is the natural home for future model-substitution, prompt-structure, and adaptive-batching layers, whose decisions couple back to filter ordering and cascade routing.

\textbf{Cross-query learning.} Both components are intra-query: state builds from the first call and is discarded at query end. Production AISQL workloads reuse predicate templates across queries, so per-predicate selectivity histories, per-template calibration curves, and distilled cost models could carry over. Persistence faces prompt and data drift, template re-binding, and isolation boundaries; a per-template warm-start cache with bounded staleness preserves the intra-query loop while removing the cold-start tax for recurring predicates. The closest classical analogue is feedback-driven optimization such as LEO~\cite{stillger2001leo}, adapted to artifacts richer than a histogram.

\textbf{Quality contracts in production.} The components' deployment criteria sit at different guarantee levels. Larch's gradient-driven refinement is best-effort: a bad prediction costs at most a few extra predicate evaluations on a row. The cascade's two thresholds control a cost-quality objective with explicit probabilistic semantics, and a sibling target-based variant (SUPG-IT~\cite{liskowski2026streamingmodelcascadessemantic}) delivers hard precision-recall guarantees at higher delegation. These form three tiers (guarantee, calibrated trade-off, best-effort) that compose loosely, so the system's guarantee is the weakest of its components. A unified contract surfacing per-component guarantees to the optimizer would let workloads swap in guarantee-preserving variants (SUPG-IT for the cascade; a budget-bounded Larch loop capping worst-case calls per row for filter ordering) and opt into a stricter overall tier. The layered framing makes such substitutions local rather than global.

\section{Conclusion}
\label{sec:conclusion}

We have developed a framework for online-learning composition at the LLM call boundary, built on the latency-window observation: a CPU-side learner's update fits inside the LLM round-trip, opening a regime classical adaptive query processing could not reach. Cortex AISQL populates three cells of the framework's design space. The two cost factors of an \texttt{AI\_FILTER} operator are decomposed across two learning components. Filter ordering reduces the expected number of LLM-evaluated predicates per row, and cascade routing reduces the fraction of evaluated predicates that escalate to the oracle. Both components instantiate the same learning pattern, with filter ordering at an online per-call cadence and cascade routing at an online per-batch cadence. Each hides its training compute behind the round-trip of the LLM call it precedes. Response caching covers the same boundary through exact memoization rather than online estimation.

A Hoare-style sequential composition rule with cost semantics shows that the two learning components compose multiplicatively under independence, yielding an analytical upper bound of $11.4\times$ total cost reduction on a representative conjunction-filter workload. Three concrete cross-component interactions, classified by the rule as two forward precondition mismatches and one backward supervision flow, lower the realistic figure to roughly $8\times$. Each is structural rather than fundamental and admits a remedy at the optimizer level.

LLM-native data systems will be defined less by the semantics of any individual operator and more by the layered, latency-budget-aware optimization architecture in which the operators are embedded. The learning pattern is one expression of that architecture. The natural next steps are empirical validation of the composition argument at scale, cross-query learning, joint controllers across components, and future layers in the design space.


\begin{thebibliography}{10}\itemsep=1pt\begin{small}

\bibitem{liskowski2025aisql}
Pawe{\l} Liskowski, Benjamin Han, Paritosh Aggarwal, Bowei Chen, Boxin Jiang,
  Nitish Jindal, Zihan Li, Aaron Lin, Kyle Schmaus, Jay Tayade, Weicheng Zhao,
  Anupam Datta, Nathan Wiegand, and Dimitris Tsirogiannis.
\newblock Cortex {AISQL}: A production {SQL} engine for unstructured data.
\newblock {\em arXiv preprint arXiv:2511.07663}, 2025.

\bibitem{HeSethi2025BigQueryAI}
Jian He and Vaibhav Sethi.
\newblock {SQL} reimagined for the {AI} era with {BigQuery} {AI} functions.
\newblock Google Cloud Blog, Data Analytics, nov 2025.
\newblock
  \url{https://cloud.google.com/blog/products/data-analytics/sql-reimagined-for-the-ai-era-with-bigquery-ai-functions}.

\bibitem{Microsoft2026FabricAIFunctions}
{Microsoft}.
\newblock Use {AI} functions (preview).
\newblock Microsoft Learn, Fabric Data Warehouse Documentation, mar 2026.
\newblock
  \url{https://learn.microsoft.com/en-us/fabric/data-warehouse/ai-functions}.

\bibitem{lin2026semasqltraditionalrelationalquerying}
Yin Lin, Tianjing Zeng, Zhongjun Ding, Rong Zhu, Bolin Ding, H.~V. Jagadish,
  and Jingren Zhou.
\newblock {SEMA-SQL}: Beyond traditional relational querying with large
  language models.
\newblock {\em arXiv preprint arXiv:2604.23477}, 2026.

\bibitem{patel2025semantic}
Liana Patel, Siddharth Jha, Parth Asawa, Melissa Pan, Carlos Guestrin, and
  Matei Zaharia.
\newblock Semantic operators: A declarative model for rich, {AI}-based analytics
  over text data.
\newblock {\em arXiv preprint arXiv:2407.11418}, 2024.

\bibitem{liu2025palimpzest}
Chunwei Liu, Matthew Russo, Michael Cafarella, Lei Cao, Peter~Baile Chen, Zui
  Chen, Michael Franklin, Tim Kraska, Samuel Madden, Rana Shahout, and Gerardo
  Vitagliano.
\newblock Palimpzest: Optimizing {AI}-powered analytics with declarative query
  processing.
\newblock In {\em Conference on Innovative Data Systems Research (CIDR)}, 2025.

\bibitem{sun2025quest}
Zhaoze Sun, Qiyan Deng, Chengliang Chai, Kaisen Jin, Xinyu Guo, Han Han,
  Ye~Yuan, Guoren Wang, and Lei Cao.
\newblock Quest: Query optimization in unstructured document analysis.
\newblock {\em arXiv preprint arXiv:2507.06515}, 2025.

\bibitem{Thalamusdb}
Saehan Jo and Immanuel Trummer.
\newblock {ThalamusDB}: Approximate query processing on multi-modal data.
\newblock {\em Proc. ACM Manag. Data}, 2(3), May 2024.

\bibitem{zhao2024hybridqueryingrelationaldatabases}
Fuheng Zhao, Divyakant Agrawal, and Amr~El Abbadi.
\newblock Hybrid querying over relational databases and large language models.
\newblock In {\em Conference on Innovative Data Systems Research (CIDR)}, 2025.

\bibitem{fu2024serverlessllm}
Yao Fu, Leyang Xue, Yeqi Huang, Andrei-Octavian Brabete, Dmitrii Ustiugov,
  Yuvraj Patel, and Luo Mai.
\newblock {ServerlessLLM}: Low-latency serverless inference for large language
  models.
\newblock In {\em 18th USENIX Symposium on Operating Systems Design and
  Implementation (OSDI 24)}, pages 135--153, 2024.

\bibitem{chung2026proxy}
Yeounoh Chung, Rushabh Desai, Jian He, Yu~Xiao, Thibaud Hottelier, Yves-Laurent
  Kom~Samo, Pushkar Kadilkar, Xianshun Chen, Sam Idicula, Fatma {\"O}zcan, Alon
  Halevy, and Yannis Papakonstantinou.
\newblock 100x cost \& latency reduction: Performance analysis of {AI} query
  approximation using lightweight proxy models.
\newblock In {\em Proceedings of the 2026 ACM SIGMOD International Conference
  on Management of Data}, 2026.

\bibitem{dai2024uqe}
Hanjun Dai, Bethany~Yixin Wang, Xingchen Wan, Bo~Dai, Sherry Yang, Azade Nova,
  Pengcheng Yin, Phitchaya~Mangpo Phothilimthana, Charles Sutton, and Dale
  Schuurmans.
\newblock {UQE}: A query engine for unstructured databases.
\newblock In {\em Advances in Neural Information Processing Systems}, 2024.

\bibitem{zhao2025accesspathsefficientordering}
Fuheng Zhao, Jiayue Chen, Yiming Pan, Tahseen Rabbani, Sohaib, Divyakant
  Agrawal, Amr~El Abbadi, Paritosh Aggarwal, Anupam Datta, and Dimitris
  Tsirogiannis.
\newblock Access paths for efficient ordering with large language models.
\newblock {\em arXiv preprint arXiv:2509.00303}, 2025.

\bibitem{Shankaretal2024}
Shreya Shankar, Tristan Chambers, Tarak Shah, Aditya~G. Parameswaran, and
  Eugene Wu.
\newblock {DocETL}: Agentic query rewriting and evaluation for complex document
  processing.
\newblock {\em Proceedings of the VLDB Endowment}, 18(9):3035--3048, 2025.

\bibitem{stillger2001leo}
Michael Stillger, Guy~M. Lohman, Volker Markl, and Mokhtar Kandil.
\newblock {LEO} -- {DB2}'s {L}earning {O}ptimizer.
\newblock In {\em Proceedings of the 27th International Conference on Very
  Large Data Bases (VLDB)}, pages 19--28, 2001.

\bibitem{avnur2000eddies}
Ron Avnur and Joseph~M. Hellerstein.
\newblock Eddies: Continuously adaptive query processing.
\newblock In {\em Proceedings of the 2000 ACM SIGMOD International Conference
  on Management of Data}, pages 261--272. ACM, 2000.

\bibitem{justen2024polar}
David Justen, Daniel Ritter, Campbell Fraser, Andrew Lamb, Allison Lee, Thomas
  Bodner, Mhd~Yamen Haddad, Steffen Zeuch, Volker Markl, and Matthias Boehm.
\newblock {POLAR}: Adaptive and non-invasive join order selection via plans of
  least resistance.
\newblock {\em Proceedings of the VLDB Endowment}, 17(6):1350--1363, 2024.

\bibitem{kabra1998efficient}
Navin Kabra and David~J. DeWitt.
\newblock Efficient mid-query re-optimization of sub-optimal query execution
  plans.
\newblock In {\em Proceedings of the 1998 ACM SIGMOD International Conference
  on Management of Data}, pages 106--117. ACM, 1998.

\bibitem{markl2004robust}
Volker Markl, Vijayshankar Raman, David Simmen, Guy Lohman, Hamid Pirahesh, and
  Miso Cilimdzic.
\newblock Robust query processing through progressive optimization.
\newblock In {\em Proceedings of the 2004 ACM SIGMOD International Conference
  on Management of Data}, pages 659--670. ACM, 2004.

\bibitem{liskowski2026larch}
Fuheng Zhao, Pawe{\l} Liskowski, Zihan Li, Benjamin Han, Puxuan Yu, Varich
  Boonsanong, Dimitris Tsirogiannis, and Anupam Datta.
\newblock {Larch}: Learned query optimization for semantic predicates.
\newblock {\em arXiv preprint arXiv:2606.07923}, 2026.

\bibitem{liskowski2026streamingmodelcascadessemantic}
Pawe{\l} Liskowski and Kyle Schmaus.
\newblock Streaming model cascades for semantic {SQL}.
\newblock {\em arXiv preprint arXiv:2604.00660}, 2026.

\bibitem{krishnamurthy1986optimization}
Ravi Krishnamurthy, Haran Boral, and Carlo Zaniolo.
\newblock Optimization of nonrecursive queries.
\newblock In {\em Proceedings of the 12th International Conference on Very
  Large Data Bases (VLDB)}, pages 128--137, 1986.

\bibitem{ibaraki1984optimal}
Toshihide Ibaraki and Tiko Kameda.
\newblock On the optimal nesting order for computing {N}-relational joins.
\newblock {\em ACM Transactions on Database Systems}, 9(3):482--502, 1984.

\bibitem{hastie1986generalized}
Trevor Hastie and Robert Tibshirani.
\newblock Generalized additive models.
\newblock {\em Statistical Science}, 1(3):297--310, 1986.

\bibitem{wood2017generalized}
Simon~N. Wood.
\newblock {\em Generalized Additive Models: An Introduction with {R}}.
\newblock Chapman and Hall/CRC, 2nd edition, 2017.

\bibitem{guo2017calibration}
Chuan Guo, Geoff Pleiss, Yu~Sun, and Kilian~Q. Weinberger.
\newblock On calibration of modern neural networks.
\newblock In {\em International Conference on Machine Learning (ICML)}, pages
  1321--1330, 2017.

\bibitem{kang2020approximate}
Daniel Kang, Edward Gan, Peter Bailis, Tatsunori Hashimoto, and Matei Zaharia.
\newblock Approximate selection with guarantees using proxies.
\newblock {\em Proceedings of the VLDB Endowment}, 13(11):1990--2003, 2020.

\bibitem{hoare1969axiomatic}
C.~A.~R. Hoare.
\newblock An axiomatic basis for computer programming.
\newblock {\em Communications of the ACM}, 12(10):576--580, 1969.

\bibitem{kaminski2018weakest}
Benjamin~Lucien Kaminski, Joost-Pieter Katoen, Christoph Matheja, and Federico
  Olmedo.
\newblock Weakest precondition reasoning for expected runtimes of randomized
  algorithms.
\newblock {\em Journal of the ACM}, 65(5):30:1--30:68, 2018.

\bibitem{chen2023frugalgpt}
Lingjiao Chen, Matei Zaharia, and James Zou.
\newblock {FrugalGPT}: How to use large language models while reducing cost and
  improving performance.
\newblock {\em arXiv preprint arXiv:2305.05176}, 2023.

\end{small}\end{thebibliography}
\end{document}